\documentclass{ws-procs975x65}
\begin{document}

\title{ISOLATED HORIZONS IN NUMERICAL RELATIVITY:\\ CONSTRUCTING THE EXCISED KERR SPACETIME\\ IN DIRAC GAUGE}

\author{NICOLAS VASSET$^*$ and J\'{E}R\^{O}ME NOVAK}

\address{Laboratoire Univers et Th\'{e}ories (LUTh), Observatoire de Paris,\\
CNRS, Universit\'{e} Paris Diderot, 5 place Jules Janssen, \\
F-92190 Meudon, France\\
$^*$E-mail: nicolas.vasset@obspm.fr}

\author{JOS\'{E} LUIS JARAMILLO}

\address{Instituto de Astrof\'{i}sica de Andaluc\'{i}a, CSIC\\
Apartado Postal 3004, Granada 18080, Spain}

\begin{abstract}
  Using a constrained formalism for Einstein equations in Dirac gauge, we propose to compute excised quasistationary initial data for black hole spacetimes in full general relativity. Vacuum spacetime settings are numerically constructed by using the isolated horizon formalism; we especially tackle the conformal metric part of our equations, assuming global stationarity. We show that a no-boundary treatment can be used on the horizon for the equation related to the conformal metric. We relate this finding to previous suggestions in the literature, and use our results to assess the widely used conformally flat approximation for computing black hole initial data. 
\end{abstract}

\bodymatter

\section{Describing stationary black hole data with isolated horizons}

 Aiming at the computation of black hole initial data in numerical relativity, this work suggests the use of an excision approach, by solving Einstein equations outside two-spheres assumed to encircle black hole singularities. In the spirit of the {\it membrane paradigm}\cite{TPM86}, we  describe black holes as physical objects characterized by their horizons. We shall use here the local characterization for black hole {\it isolated horizons}, associated to a nonevolving black hole region. Isolated horizons are defined \cite{AshBF} as three-dimensional null tubes foliated by marginally trapped surfaces (geometrical expansion along the outer future null normal vanishes), with the additional restriction that the extrinsic geometry of the tube is not evolving along the null generators. Those features will translate in our computation as boundary conditions on our chosen (3+1) variables associated to the metric\cite{CookPf04, JAL07}.

We choose to write Einstein equations following a (3+1) fully constrained formalism\cite{Bon04}, with maximal slicing and spatial Dirac gauge. In the vacuum, asymptotically flat, single black hole spacetime setting, we end up with a subset of three elliptic equations (two scalar, one vector) to solve, associated to the classical lapse $N$, shift $\beta^{i}$ and conformal factor $\psi$ (3+1) variables, and expressing the Einstein constraints as well as the gauge choice. Those equations require boundary conditions at the inner excised surface, provided by the prescription of this surface to correspond to a slice of an isolated horizon. Two free parameters allow to fix the mass and angular momentum of the black hole. Solving for those three PDE on a spacelike slice, alongside with setting the conformal 3-metric $\tilde{\gamma}^{ij}= \psi^{4} \gamma^{ij}$ ($\psi$ being the conformal factor) to be the usual flat metric leads to vacuum black hole initial data with the classical conformal flatness hypothesis. \cite{CookPf04}

\section{Solving for the conformal metric: the no-boundary approach}
 
  Using a global stationarity hypothesis, our formalism provides a fourth elliptic-like equation for the deviation tensor $h^{ij}$, being the non-flat part of the conformal metric ($h^{ij} = \tilde{\gamma}^{ij}- f^{ij}$, with $f^{ij}$ the reference flat metric)\cite{Bon04}:
\begin{equation}
\Delta h^{ij} -\frac {\psi^{4}}{N^{2}} \mathcal{L}_{\beta}
\mathcal{L}_{\beta}h^{ij} = \mathcal{S}^{ij}(h^{ij}, N, \psi, \beta). \nonumber
\end{equation}
with $\mathcal{L}$ the classical Lie derivative. Going beyond the conformal flatness assumption requires to solve this equation, and particularly to treat the boundary condition problem. Concerning boundary conditions for the conformal metric in excised initial data, a few proposals have been formulated in the literature. Jaramillo\cite{Jar09} proposed to express a geometrical property of the isolated horizon, namely the invariance of the extrinsic geometry. Cook and Baumgarte\cite{CookBaum08} expressed that this boundary value could be arbitrarily prescribed with no effect on the physical content of the data. On the contrary, we describe in this work a {\it no-boundary treatment}, arguing that the operator is weakly singular at the horizon, and thus can be inverted without additional information at the boundary. The data obtained are coherent with the prescription of Jaramillo, but not with the one of Cook and Baumgarte. Details on the resolution of this tensor equation, including an extraction of the two relevant scalar degrees of freedom in Dirac gauge\cite{JNJLCNV}, can be retrieved from the article related to this report\cite{NVJNJLJ}. Let us note that this result has been obtained one more time since this study\cite{Grand10}, with a different approach in the treatment of the conformal metric.  

\section{Results and pertinence of the study}

\begin{figure}
\begin{center}
\psfig{file=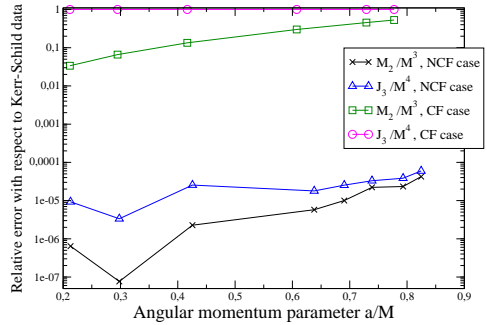,width=2.1in, angle=-90}
\end{center}
\caption{Computations of the rescaled second-order mass multipole $(
    {M_{2}}/{M^3})$ ($M$ being the ADM mass computed at infinity) and the rescaled third-order angular multipole $({J_{3}}/{M^{4}})$. Relative differences with respect
    to values for a
  Kerr-Schild analytical horizon are displayed in the conformally flat (CF) or the non-conformally flat (NCF) setting, against the usual (dimensionless) rotation parameter for a Kerr spacetime.}
\label{fig1.eps}
\end{figure}

The data obtained from our computations verify the Einstein constraints, as well as the (3+1) evolution equation in stationarity, so that we are able to reconstruct the whole spacetime by simple time translation. The initial data settings (single black hole in vacuum spacetime) suggest that we should obtain from this study the Kerr solution in Dirac gauge (which has no analytic formulation). We test this assumption with several markers: first, although computations are performed with no spatial symmetry hypothesis, the final data are indeed globally axisymmetric. Computation of the virial integral at infinity\cite{Virial94} shows that contrary to the corresponding conformally flat data, our spacetime appears to be stationary. Finally, performing a source multipole decomposition\cite{AshENPVDB} of the isolated horizon slice, and comparing with values for an horizon of Kerr geometry constructed analytically, we show that our horizon geometry is the one of a Kerr black hole to a very reasonable accuracy (see Fig. \ref{fig1.eps}). Equating the isolated horizon multipoles suffices to identify both spacetimes\cite{AshENPVDB}. The corresponding data with a conformal flatness hypothesis show however a clear discrepancy.

To our knowledge it is the first time the full 3-metric (including the conformal part) for black hole data is obtained numerically using only a stationarity assumption. Further numerical work using this approach to binary black hole settings could lead to improvements in the computation of initial data for such systems. 

\section*{Acknowledgements}
N.V and J.N were supported by the A.N.R. grant No. BLAN07-1-201699 entitled ``LISA Science''. J.L.J. has been supported by the Spanish MICINN under Project No. FIS2008-06078-C03-01/FIS, and Junta de Andaluc\'{i}a under projects No. FQM2288 and No. FQM219.

\bibliographystyle{ws-procs975x65}
\bibliography{ws-pro-sample}

\begin{thebibliography}{8}

\bibitem{TPM86} K.~Thorne, R.~Price and D.~MacDonald, {\it Black Holes: The
  membrane paradigm}, Yale University Press (1986). 

\bibitem{AshBF} A.~Ashtekar, C.~Beetle and S.~Fairhurst, Class.\ Quantum
  Grav.\ {\bf 16}, L1--L7 (1999).

\bibitem{CookPf04} G.~B.~Cook and H.~Pfeiffer, Phys.\ Rev.\ D {\bf 70}, 104016 (2004).

\bibitem{JAL07} J.~L.~Jaramillo, M.~Ansorg and F.~Limousin, Phys.\ Rev.\ D
  {\bf 74}, 087502 (2007).

\bibitem{Bon04} S.~Bonazzola,  E.~Gourgoulhon,  P.~Grandcl\'ement and
  J.~Novak, Phys.\ Rev.\ D {\bf 70}, 104007 (2004). 

\bibitem{Jar09} J.~L.~Jaramillo, Phys.\ Rev.\ D {\bf 79}, 087506 (2009).

\bibitem{CookBaum08} G.~B.~Cook and T.~Baumgarte, Phys.\ Rev.\ D {\bf 77},
  064023 (2008).


\bibitem{JNJLCNV} J.~Novak, J.~L.~Cornou, N.~Vasset, J.~Comp.~Phys. {\bf 229}, 399-414 (2010)

\bibitem{NVJNJLJ} N.~Vasset, J.~Novak, J.~L.~Jaramillo, Phys.~Rev.~D. {\bf 79}, 124010 (2009)

\bibitem{Grand10} P.~Grandcl\'{e}ment, J.~Comp.~Phys, in press.(2010)

\bibitem{Virial94} E.~Gourgoulhon and S.~Bonazzola, Class.\ Quantum Grav.\
  {\bf 11}, 443--452 (1994).

\bibitem{AshENPVDB} A.~Ashtekar, J.~Engle, T.~Pawlowski and C.~Van Den
  Broecke, Class.\ Quantum Grav.\ {\bf 21}, 2549--2570 (2004). 

\end{thebibliography}

\end{document}